\documentclass[preprint,numberedappendix]{aastex}
\usepackage{color}
\usepackage{longtable}
\newcommand{\msun}{M$_{\odot}$}
\newcommand{\lsun}{L$_{\odot}$}
\newcommand{\um}{$\mu$m}

\newcommand{\gs}{{\tt getsources}}%

\shorttitle{Massive Cold Clumps in NGC~7538}
\shortauthors{Fallscheer et al.}

\begin{document}


\title{\emph{Herschel}\footnote{\emph{Herschel} 
is an ESA space observatory that has science instruments provided by 
European-led Principal Investigator consortia with important 
participation from NASA.} ~Reveals Massive Cold Clumps in NGC~7538}


\author{C.~Fallscheer\altaffilmark{1,2}, 
M.~A.~Reid\altaffilmark{3},
J.~Di~Francesco\altaffilmark{2,1},
P.~G.~Martin\altaffilmark{4}, %added CITA
M.~Hennemann\altaffilmark{5},
T.~Hill\altaffilmark{5},
Q. Nguyen-Luong\altaffilmark{4,5}, %added CITA
F.~Motte\altaffilmark{5},
A.~Men'shchikov\altaffilmark{5},
Ph.~Andr\'e\altaffilmark{5},
D.~Ward-Thompson\altaffilmark{6},
M.~Griffin\altaffilmark{7},
J.~Kirk\altaffilmark{6},
V.~Konyves\altaffilmark{5,8},
K.~L.~J.~Rygl\altaffilmark{9}, %new addition
M.~Sauvage\altaffilmark{5},
N.~Schneider\altaffilmark{10}, %changed affil.
L.~D.~Anderson\altaffilmark{11},
M.~Benedettini\altaffilmark{9}, %new addition
J.-P.~Bernard\altaffilmark{12,13}, %new addition
S.~Bontemps\altaffilmark{10},
A.~Ginsburg\altaffilmark{14},
S.~Molinari\altaffilmark{9},
D.~Polychroni\altaffilmark{15}, %new addition
A.~Rivera-Ingraham\altaffilmark{12,13}, %new addition
H.~Roussel\altaffilmark{16},
L.~Testi\altaffilmark{17},
G.~White\altaffilmark{18,19},
J.~P.~Williams\altaffilmark{20},
C.~D.~Wilson\altaffilmark{21},
M.~Wong\altaffilmark{22,2},
A.~Zavagno\altaffilmark{11}
}
\altaffiltext{1}{Department of Physics \& Astronomy, University of Victoria, PO Box 355, STN CSC, Victoria, BC, V8W 3P6, Canada}
\altaffiltext{2}{National Research Council Canada, 5071 West Saanich Road, Victoria, BC V9E 2E7, Canada}
\altaffiltext{3}{Department of Astronomy and Astrophysics, University of Toronto, Toronto, ON M5S 3H4, Canada}
\altaffiltext{4}{Canadian Institute for Theoretical Astrophysics, University of Toronto, Toronto, ON M5S 3H8, Canada}
\altaffiltext{5}{Laboratoire AIM, CEA/DSM-CNRS-Universit Paris Diderot, 
IRFU/Service dÕAstrophysique, Saclay, 91191 Gif-sur-Yvette, France}
\altaffiltext{6}{Jeremiah Horrocks Institute, University of Central Lancashire, Preston, Lancashire, PR1 2HE, UK }
\altaffiltext{7}{School of Physics and Astronomy, Cardiff University, Queen's Buildings, The Parade, Cardiff CF24 3AA, UK}
\altaffiltext{8}{IAS, CNRS (UMR 8617), Universit\'{e} Paris-Sud 11, B\^{a}timent 121, 91400 Orsay, France}
\altaffiltext{9}{INAF Ð Istituto di Astrofisica e Planetologia Spaziali, Area di  
Ricerca di Tor Vergata, via Fosso del Cavaliere 100, 00133 Roma, Italy}
%\altaffiltext{8}{Istituto di Astrofisica e Planetologia Spaziali (IAPS-INAF), via del
%Fosso del Cavaliere 100, 00133 Roma, Italy}
%\altaffiltext{10}{Universit\'{e} de Toulouse, UPS, CESR, 9 av. du colonel Roche, 31028, Toulouse, France}
\altaffiltext{10}{Universit\'{e} de Bordeaux, LAB, UMR 5804, F-33270, Floirac, France}
\altaffiltext{11}{Laboratoire d'Astrophysique de Marseille, CNRS/INSU – Universit\'e de Provence, 13388 Marseille Cedex 13, France}
\altaffiltext{14}{University of Colorado, Boulder, CO, USA}
\altaffiltext{15}{University of Athens, Faculty of Physics, Department of Astrophysics, Astronomy and Mechanics, Panepistimiopolis, 15784 Zografos, Athens, Greece}
%\altaffiltext{10}{Istituto di Fisica dello Spazio Interplanetario, INAF, via del 
%Fosso del Cavaliere 100, 00133 Roma, Italy}
\altaffiltext{12}{Universit\'{e} de Toulouse, UPS-OMP, IRAP, F-31028 Toulouse Cedex 4, France}
\altaffiltext{13}{CNRS, IRAP, 9 Av. colonel Roche, BP 44346, F-31028 Toulouse Cedex 4, France}
\altaffiltext{16}{Institut d'Astrophysique de Paris, Universit\'e Pierre \& Marie Curie, 98bis boulevard Arago, 75014 Paris, France}
\altaffiltext{17}{ESO, Karl Schwarzschild-Strasse 2, 85748 Garching bei M\"{u}nchen, Germany}
\altaffiltext{20}{Institute for Astronomy, University of Hawaii, 2680 Woodlawn Drive, Honolulu, HI 96822, USA}
\altaffiltext{18}{The Open University, Department of Physics and 
Astronomy, Milton Keynes MK7 6AA, UK}
\altaffiltext{19}{The Rutherford Appleton Laboratory, Chilton, Didcot, Oxfordshire OX11 0NL}
\altaffiltext{21}{Dept. of Physics \& Astronomy, McMaster University, Hamilton, Ontario, L8S 4M1, Canada}
\altaffiltext{22}{Department of Physics and Astronomy, University of British Columbia, Vancouver, BC V6T 1Z1, Canada}


\begin{abstract} 

We present the first overview of the \emph{Herschel} observations of the nearby 
high-mass star-forming region NGC~7538, taken as part of the \emph{Herschel} 
imaging study of OB Young Stellar objects (HOBYS) Key Programme.  These 
PACS and SPIRE maps cover an approximate area of one square degree at 
five submillimeter and far-infrared wavebands.  We have 
identified 780 dense sources and classified 224 of those.  
 With the intention of investigating the existence of cold massive starless or class 0-like clumps that would have the potential to form intermediate- to high-mass stars, we further isolate 13 %instances 
clumps as the most likely candidates for followup studies. 
%We further isolate and investigate 17 %42 
%instances of candidate cold massive starless clumps.  These possible sites of the early stages of high-mass star formation 
%as candidate sites of the early stages of 
%the existence of cold massive starless clumps able to form massive stars
%among which 
%we isolate 17 as candidate sites of the early stages of high-mass star 
%formation.  We investigate these 
%These latter sources 
These 13 clumps %sites
have masses in excess of %20
40~\msun~and 
temperatures below 15~K.   They range in size from 0.4~pc to 2.5~pc and have densities between $3\times10^{3}$~cm$^{-3}$ to $4\times10^{4}$~cm$^{-3}$. Spectral energy distributions are then used to 
characterize their energetics and evolutionary state through a luminosity-mass diagram.
%We construct spectral energy 
%distributions (SEDs) for these objects and thereby characterize their 
%energetics on a luminosity-mass diagram.  
%Our maps reveal that the 
NGC~7538 
%region 
has a highly filamentary structure, previously unseen in the dust 
continuum of existing submillimeter surveys.  We report the most complete imaging to date of a large, 
evacuated ring of material in NGC~7538 which is bordered by many cool sources. %condensations, 
%which may suggest that triggering is the mechanism driving star formation in this region. %suggestive of triggered star formation.
 \end{abstract}


\section{Introduction}

The European Space Agency's \emph{Herschel} Space Observatory \citep{pilb} probes the 
cold, dusty precursors of stars in unprecedented numbers and detail.  
\emph{Herschel}'s submillimeter and far-infrared wavebands span the 
peak of the spectral energy distribution (SED) where cold cores emit the bulk of their radiation. %of typical cold cores.  
The combination of \emph{Herschel}'s spectral coverage and high angular resolution 
allows, for the first time, the identification of large numbers of potential 
star-forming cores ($\sim$0.1 pc) and clumps ($\sim$0.5 pc) %\citep{beut2007} 
in high-mass star-forming regions and to measure their 
dust properties.  In this paper, we report the first \emph{Herschel} 
results on the nearby high-mass star-forming region NGC~7538 observed as part 
of the \emph{Herschel} imaging study of OB Young Stellar Objects (HOBYS, 
\citealt{hobys}) Key Programme.  At a %n estimated distance of only 2.8~kpc 
 trigonometric parallax distance of only 2.7~kpc from the solar system \citep{mos09},  \objectname{NGC 7538} is a relatively 
nearby region of high-mass star formation and an excellent place to search 
for the precursors of future high-mass stars.

%	The \objectname{NGC 7538} molecular cloud complex has been studied 
%extensively.  
Although \objectname{NGC~7538} is best known for the bright 
\ion{H}{2} region surrounding the source known as IRS~5 \citep{cgg78}, 
star-formation %activity
 is distributed widely throughout the larger molecular 
cloud.  Each of the three brightest infrared sources IRS 1, 2, and 3 has 
its own associated compact \ion{H}{2} region \citep{ww74}.  Collectively, 
these three sources are embedded in a region rich with outflows and 
photodissociation fronts \citep{davis1998}.  Observations of a 12\arcmin$\times$8\arcmin~region at 
850~\um~\citep{rw05} showed that all of these sources are embedded in an extensive 
network of filaments and compact sources comprising several thousand solar 
masses of 
gas and dust.  Elsewhere in the region, \citet{fries07} found a filamentary 
complex in an 18\arcmin$\times$18\arcmin~map of C$^{18}$O(2--1) around the cold cloud G111.80+0.58 
(at $\alpha,\delta$ (J2000) = 23:16:22, 61:22:47; G111 
hereafter).  The G111 complex contains about a dozen candidate cold, 
high-mass clumps, % cores, 
several with masses exceeding 100~\msun~\citep{fries07}.  All of these regions are labeled in Figure \ref{fig:threecol}.

	The \emph{Herschel} data presented here %significantly 
expand our spatial coverage %picture 
of this region by nearly an order of magnitude, %  96(12'x8')+324(18'x18')=420sq.arcmin vs (60'x60')3600sq.arcmin} 
revealing the full extent of the filamentary structure in 
the area as well as a large population of compact, potentially star-forming 
sources.  \emph{Herschel}'s ability to 
both detect and characterize cold, high-mass sources %like this one 
make it an especially exciting tool for the study of high-mass star 
formation in nearby regions such as NGC~7538.

\section{Observations}
\subsection{\emph{Herschel} Data}
\label{sec:data}

\begin{figure} %figure 1
\centering
%\plotone{ngc7538_3colour_70b_160g_250r_scana_v2.eps}
%\includegraphics[scale=.9]{ngc7538_3colour_70b_160g_250r_scana_v2.eps}
%\includegraphics[scale=.6]{rgb_ds9.ps}
%\includegraphics[scale=.6]{rgb_7538_v2.eps}
\includegraphics[scale=.6]{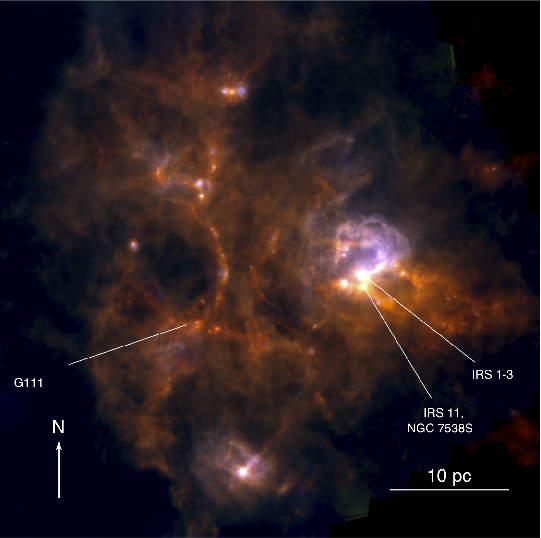}
\caption{Three-color image of an approximately 
50\arcmin $\times$ 50\arcmin~portion of NGC~7538.  %Closer to 53'x53'
The wavebands included are SPIRE 
250~\um~(\emph{red}), PACS 160~\um~(\emph{green}), and PACS 
70~\um~(\emph{blue}).  %The stretch and contrast level of each waveband 
%have been adjusted to accentuate color variations.  
The eastern half of 
the image is dominated by a prominent ring-like feature of uncertain 
origin.  %Axes are in J2000 coordinates. %Note the small saturated regions at the two brightest 
%peaks (see \S\ref{sec:data}).  
\label{fig:threecol}}
\end{figure}

The observations of a $\sim 1\degr \times 1\degr$ portion of \object{NGC 
7538} were made by \emph{Herschel} on 14 December 2009 as part of the HOBYS Key Programme.  
The data were acquired using the PACS \citep{pacs} and SPIRE \citep{spire} 
cameras working in parallel mode with a scanning speed of 20\arcsec/second (OBSIDs: 134218808, 1342188089).
Images were obtained simultaneously with PACS at 70~\um~and 160~\um~and 
SPIRE at 250~\um, 350~\um, and 500~\um.  These five wavebands range 
in angular resolution from 5.6\arcsec~at 70~\um~to 36\arcsec~at 500~\um.  
%The data from both cameras were reduced and calibrated using HIPE version 
%5.0.1975.  The maps were made using version 4.0 of  
%Scanamorphos\footnote{\textsl{http://www2.iap.fr/users/roussel/herschel/}} \citep{rous2012}.  
 The calibration and deglitching of the Level 0 PACS and SPIRE data were done using HIPE\footnote{HIPE is a joint
development software by the \emph{Herschel} Science Ground Segment Consortium,
consisting of ESA, the NASA \emph{Herschel} Science Center, and the HIFI, PACS,
and SPIRE consortia.} version 9.0.  The level 1 data were then used to produce maps with version 18 of the Scanamorphos software
package %\footnote{\it http://www2.iap.fr/users/roussel/herschel/}
\citep{rous2012}.
%This comes directly from Quang's W48 paper: 
%The beam sizes are $\sim$6\arcsec, 12\arcsec,18\arcsec, 25\arcsec, and 37\arcsec~and the final images have 1$\sigma$ rms of 0.02~Jy/1.4 \arcsec$-pixel, 0.08~Jy/2.8$\arcsec$-pixel, 1~Jy/beam, 1.1~Jy/beam, and 1.2~Jy/beam for 70\,$\micron$, 160\,$\micron$, 250\,$\micron$, 350\,$\micron$, and 500\,$\micron$, respectively.  
Figure~\ref{fig:threecol} shows a 
three-color image made from SPIRE and PACS data.  The stretch and contrast of each waveband have been manipulated slightly to accentuate 
color variations.  The bright region which 
dominates the western half of the image coincides with the void in the 
\ion{H}{2} region.  This bright complex contains the aforementioned IRS sources 
as well as the young stellar object NGC~7538S ($\alpha,\delta$ (J2000)= 23:13:45, +61:26:51).  The two brightest emission peaks in the image saturated the detectors in the SPIRE 
250~\um~band.  These peaks are coincident with IRS~1--3 (at 
$\alpha,\delta$ (J2000)= 23:13:45, +61:28:10) and IRS~11 (at $\alpha,\delta$ (J2000)= 
23:13:44, +61:26:49).  Both peaks were re-observed with \emph{Herschel} in bright source mode 
to fill in the holes in the map (OBSID: 1342239268). These observations were taken on 13 Feb 2012 and the saturated pixels were replaced in the images using the method described in Nguyen Luong et al. (subm.). Figure~\ref{fig:threecol} also highlights the highly filamentary nature of the emission.
%; for the moment, these sources have been 
%excluded from our analysis (they are, in any case, definitely not 
%pre-stellar).
%The stretch and contrast of each waveband in
%Figure~\ref{fig:threecol} has been manipulated slightly to accentuate 
%color variations.  %In reality, the emission is fairly ``grey,'' especially 
%among the SPIRE wavebands.  
%This figure also highlights the highly filamentary nature of the emission.
%To highlight the highly filamentary nature of 
%the emission, we have also applied a gentle unsharp mask to the data shown in 
%Figure~\ref{fig:threecol}.

 Using the 160~\um, 250~\um, 350~\um, and 500~\um~data, we constructed H$_{2}$ column density and dust temperature maps of the region using the IDL $\chi^{2}$ minimization fit routine \emph{mpfitfun}.  These maps are shown in Figure \ref{fig:dentemp} and include labels of all of the sources mentioned above.

\subsection{JCMT data}

% Observations of the NGC 7538 region in CO(3--2) at 345 GHz were carried out with the James Clerk Maxwell Telescope (JCMT) on 9 October 2007.  
 To obtain CO(3--2) emission observations, we observed NGC~7538 with the HARP 
instrument at the James Clerk Maxwell Telescope\footnote{The James Clerk Maxwell Telescope is operated by the Joint Astronomy Centre on behalf of the Science and Technology Facilities Council of the United Kingdom, the National Research Council of Canada, and (until 31 March 2013) the Netherlands Organisation for Scientific Research.} (JCMT). % on 9 October 2007.  
HARP is a 4 $\times$ 4 array of heterodyne receivers that
can observe 325-375~GHz.  HARP was tuned to observe CO(3--2) at 345.7959899~GHz \citep{pick98}, and the JCMT's ACSIS correlator was configured
to observe the line over a 1 GHz wide band with 2048 channels having a 
velocity resolution of 0.42~km~s$^{-1}$ channels.  The NGC~7538 region was 
divided into 36 fields each 12$^{\prime}$ $\times$ 12$^{\prime}$ in size, 
and spaced in R.A. or decl. by 10$^{\prime}$, that covered essentially the 
same region observed by \emph{Herschel}.  Each field was observed in on-the-fly 
mode using an offset position found to be free of CO(3--2) emission located at 
23$^{h}$20$^{m}$55.72$^{s}$, +60$^{\circ}$50$^{\prime}$00.4$^{\prime\prime}$ 
(J2000).  After each scan along the length of a field, the array was moved 
in either R.A. or decl. by 1/4 of its extent (i.e., $\sim$30$^{\prime\prime}$) 
between scans to obtain samples of each position by several receivers.  The 
observations were obtained throughout semesters 07B, 09B, and 10B.

The JCMT data were reduced using standard procedures within the Starlink 
package.  Integrations from each receiver were checked visually for baseline
ripples or extremely large spikes, and any such affected data were removed 
from the ensemble.  Each integration had its baseline subtracted, its frequency 
axis converted to velocities in the local standard of rest frame, and its 
outer velocities trimmed using scripts kindly supplied by T. Van Kempen.  
The data for each field were co-added into final spectral cubes, and then 
arranged into a final mosaic of all fields of about 1$^{\circ}$ $\times$
1$^{\circ}$ in extent.  The typical 1 $\sigma$ rms sensitivity reached was 
$\sim$0.6 K per channel on the $T_{A}^{*}$ scale.  Spectra were converted 
to main beam brightness temperature using an efficiency measured from planetary 
observations of 0.75 (P. Friberg, private communication).
% an assumed efficiency of 0.75 (P. 
%Friberg, private communication).

\section{Results}
\subsection{SED Fitting}
\begin{figure} %Figure 2 -- old Figure A1
\centering
\includegraphics[angle=-90,scale=.9]{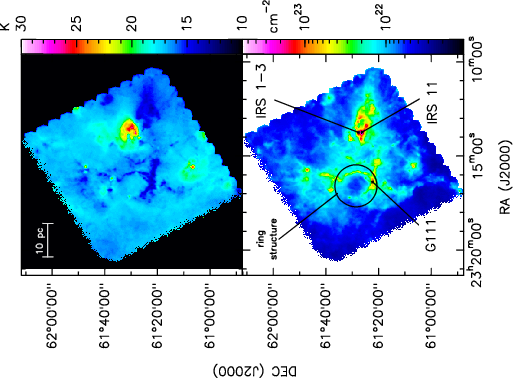}
\caption{Dust temperature (top) and molecular hydrogen column density (bottom) maps of NGC~7538.  The ring structure discussed in Section \ref{sec:ring} as well as several prominent objects in the region are labeled in the column density plot. \label{fig:dentemp}}
\end{figure}

Figure~\ref{fig:dentemp} shows the column density (N$_{H_2}$) and dust temperature maps of NGC~7538.  To produce these maps, we corrected the arbitrary zero-point flux offset for each PACS and SPIRE map with data from the IRAS and Planck telescopes \citep[see][]{bern2010}.  After applying the offsets, each PACS and SPIRE map was convolved to the 500 micron beam size (36$\arcsec$) and regridded to the same pixel resolution as the 500~\um~map.  We used the IDL routine mpfitfun to fit a modified blackbody function to each pixel, assuming the dust spectral index, $\beta$=2.0 and the dust opacity per unit mass column density, $\kappa_{\nu}$=0.1~cm$^2$/g for a reference wavelength of 300~\um.   For these fits, we exclude the 70~\um~data since the emission from this short wavelength more likely results from warmer material rather than the cold dust component traced by the longer wavelengths which we are most interested in \citep[e.g.][]{hill2011}.  For the NGC~7538 region, line-of-sight dust temperatures predominantly range between 12~K and 25~K with a mean temperature of 17~K and column densities vary from 3$\times$10$^{21}$~cm$^{-2}$ to 4$\times$10$^{23}$~cm$^{-2}$.
%The numbers in the next paragraph come from ngc7538/temp_dens/massconversion.pro and /tempdist.pro
%total mass: 393000Msun
%mass with >10^22 cm-2 colden: 214000Msun
%
 From the column density map, we calculate a total mass of the region to be nearly 4$\times$10$^{5}$~\msun~which is in agreement with \citet{unge2000}.  Over half of this mass is contained in high-column density structures ($>$10$^{22}$~cm$^{-2}$).

%\section{A Ring of Triggered Star Formation?}
\subsection{The Ring Structure} \label{sec:ring}

\begin{figure} %Figure 3 -- old Figure 5
\begin{center}
%\includegraphics[width=6.0in,angle=270]{ngc7538_ring_cont_v16_2by3.eps} 
%Figure produced with ngc7538/figures/ring_structure/plot_ring.greg
%Original figure is ngc7538/figures/ring_structure/ring_plot.eps
%\includegraphics[width=6.0in,angle=270]{ring_plot.eps} 
\includegraphics[width=6.0in,angle=270]{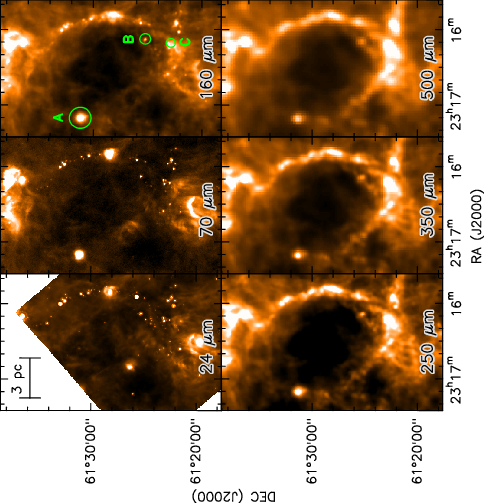} 
\caption{Images of the ring structure from \emph{Spitzer}'s MIPS (24~\um) camera, as well as \emph{Herschel}'s PACS (70~\um~and 
160~\um) and SPIRE (250~\um,~350~\um,~and 500~\um) instruments.  The sources circled in the 160~\um~image are discussed further in Section \ref{sec:getsources}. \label{fig:ring}}

%The 
%SEDs of the sources circled in the 160~\um~image are shown in 
%Figure \ref{fig:sedfig}.  \label{fig:ring}}  %G111 is the extended bright region which is 
%prominently visible toward the bottom of each panel, indicated by a plus (+) 
%symbol in the 350~\um~panel. \label{fig:ring}} 
\end{center}
\end{figure}

\begin{figure} %Figure 4 --old Figure 6
\begin{center}
\includegraphics[width=4.5in,angle=270]{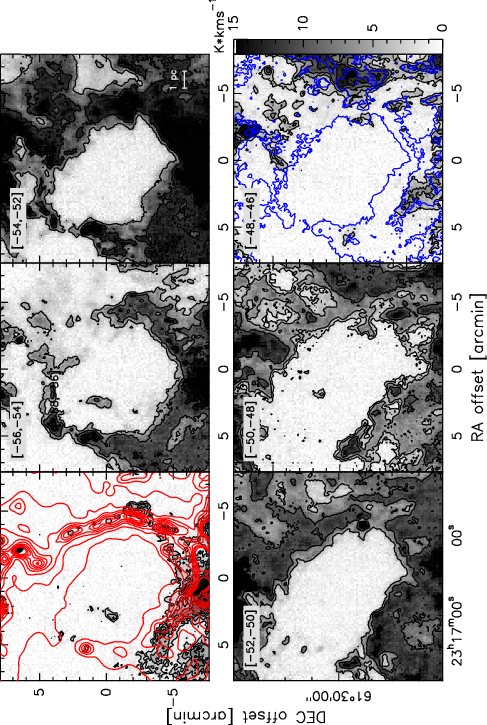} 
\caption{Integrated intensity maps of the ring %bubble 
region in JCMT CO(3--2) data.  The central pixel is at RA=23h16m26.5s decl.=61$^{\circ}$29'56'' (J2000). For each panel, contours start at 1$\sigma$ and continue in steps of 1$\sigma$ {\textbf (with the exception of the [-50,-48] channel which starts at 2 $\sigma$ and increases in steps of 2 $\sigma$)} where $\sigma$ is 2.0, 4.5, 5.9, 4.8, 4.9, and 2.9~K~km~s$^{-1}$ for the panels in order of increasing velocity.   The red contours in the upper left panel are the column density from 5$\times$10$^{21}$~cm$^{-2}$ to 5$\times$10$^{22}$~cm$^{-2}$ in steps of 5$\times$10$^{21}$~cm$^{-2}$. The blue contour lines in the bottom right panel are for the ring %bubble 
region integrated over the entire velocity range (-58~km~s$^{-1}$ to -41~km~s$^{-1}$) of emission. These contours start at 1 $\sigma$ and continue in steps of 1.5 $\sigma$ where $\sigma$=16.0~K~km~s$^{-1}$.  These intensities are on the $T_{A}^{*}$ scale. \label{fig:jcmt}} 
\end{center}
\end{figure}

	A striking feature of the NGC~7538 maps is the existence %preponderance 
of ring-like 
features, especially the nearly complete ring which dominates the eastern 
section of the \emph{Herschel} map (see Figure \ref{fig:threecol}).  A zoomed-in view of this region is shown in Figs.~\ref{fig:ring} and \ref{fig:jcmt}. %half of the frame (see Figure \ref{fig:threecol}).  
Portions of this ring 
were previously detected by \citet{fries07} in C$^{18}$O(2--1) emission and 
by \citet{fries08} in the \emph{Spitzer} continuum bands.  Here, we also present JCMT CO(3--2) data which include the entire ring structure.  Such a large, 
well-defined ring has not been seen in the other regions observed in the HOBYS survey, e.g., Rosette \citep{schn2010}, 
W48 \citep{luong}, or Cygnus X \citep{henn2012}, %Vela C \citep{hill2011}, or M16 \citep{hill2012}, 
emphasizing the peculiarity of this feature\footnote{Note that small, less prominent bubbles were observed by \citet{zava2010} in RCW120 and by \citet{ande2012}, and that \citet{chur2006} provide a catalog of bubbles seen with \emph{Spitzer} in the Galactic plane.  However, as we do not see evidence for spherical symmetry in NGC~7538, the ring structure may be a different type of object.  %\citet{mini2013} find a much smaller bubble in Vela C, and 
Some less clean rings are observed  in M16 \citep{hill2012} and W3 \citep{rive2013}.}.  
%The lack of similar objects in these other high mass star formation regions emphasizes the peculiarity of this feature. 

As shown in Figure~\ref{fig:ring}, the ring is prominent in thermal dust emission at wavelengths of
160~\um~and longer.  It is visible as a string of point sources 
in the PACS 70~\um~image and the \emph{Spitzer} MIPS 24~\um~image (see 
Figure \ref{fig:ring}).  The ring is a nearly complete ellipse with major and 
minor axes of approximately 10.6~pc and 7.4~pc, respectively.  The ring may be the edges of
%has the appearance of 
a bubble produced by an internal energetic source, but its 
origin is not yet clear.  According to an exhaustive catalog of both known 
and candidate O and B stars \citep{reed}\footnote{Vizier catalog V/125}, no 
such stars lie within the ring.  Although there is an A0 type star within the ring, the nearest known massive star is a B star 
about 4~pc east of the ring.  A search of archival data has so far revealed 
no MSX, IRAS, or radio continuum sources within the ring which might 
account for its existence.

The JCMT CO(3--2) data also show no indication of objects present that may be responsible for forming the ring.  In the channel map presented in Figure \ref{fig:jcmt}, the emission from the channels spanning the ring's velocity range is shown.  %Even 
At this longer wavelength (870~\um), a driving source is still not evident.  

Aside from a small feature most prominent in the [-56,-54]~km~s$^{-1}$ channel, the region within the ring appears completely devoid of material in its entire 17~km~s$^{-1}$ velocity range.  A spherically symmetric object such as a bubble would likely exhibit emission within the ring at the higher velocities in the channel map, which we do not see.  %One plausible scenario is that we are 
It is more likely that we are observing a bubble that has broken out of the molecular cloud, producing a ring of CO emission (similar to smaller molecular rings seen by \citet{beau2010} around the Churchwell Spitzer bubbles.)
%It may be that a bubble's higher velocity components within the outer edges are optically thin and not detected by the JCMT data, but this is speculative.}  
Such a void (with a typical column density of 6$\times$10$^{22}$~cm$^{-2}$) is in stark contrast to the rest of the NGC~7538 region as well as the other HOBYS fields which are dominated by low levels of diffuse emission (typically higher than 10$^{22}$~cm$^{-2}$) throughout the maps. %nearly everywhere.  
This lack of diffuse continuum {\emph and} line emission indicates that the ring was created through different mechanisms than %in a different fashion from
 the methods by which other filaments in these star formation regions formed.  Of course, it is possible that the ring could be a coincidental alignment of curved filaments, but the lack of diffuse gas within the ring would suggest otherwise.

While the ring appears elliptical in the \emph{Herschel} data and at the redder velocities in the CO data, it becomes even more elongated in the northeast-southwest direction at the bluer CO velocities.  The integrated intensity contours overplotted in the bottom right panel of Figure \ref{fig:jcmt} retain this dominant oblong shape.  Although the ring is brightest at the southern end closest to G111, these integrated intensity contours establish the nearly-closed morphology of this ring structure.

\subsection{Source Extraction}\label{sec:getsources}

	To identify compact sources in the field, we used version 1.120916 %1.110722 
of the \gs~algorithm \citep{gs1,gs2}.  Since the PACS and SPIRE instruments are so %intrinsically 
sensitive, we detect significant extended, diffuse emission which makes source 
extraction more %from the images 
difficult.  Rather than subtract a global background from the 
entire image, as is sometimes done, \gs~separates emission across a wide 
range of angular scales and then uses this information to identify sources 
as peaks relative to their local backgrounds.  Thus, compact sources 
are extracted without imposing any parametrization on either their 
structure or that of the diffuse background.  The algorithm assigns, to each 
source identified in each waveband, a shape, size, peak flux, total flux and significance value.  %The different resolution at each wavelength creates the potential for
Note that it is possible for sources detected at longer wavelengths to be resolved into multiple objects at shorter wavelengths due to increasing resolution.  %This potential for multiplicity is suppressed, however, by passing 
 However, \gs passes information on from the higher resolution images %when intermediate source lists are determined for the 
to the extractions of lower resolution images.  For each object detected in the higher resolution maps, the final catalog assigns a single non-overlapping flux value at all the wavelengths for which a significant detection is made.  The significance parameter is analogous to a 
signal-to-noise value in a single waveband.  %Quang's 2011 paper 535,76 paper section 4.1 has a good description of this

	Using this method, we have preliminarily identified 780 %approximately 800
	%a total of 
	%{\color{red}780 = number of sources in semif.cat.}%817 
compact sources in 
NGC~7538.  Of these, many were detected in fewer than the five possible wavebands.  In such cases, the subset of wavebands in 
which the sources were detected varied.  A common pattern was that
sources were detected with SPIRE, but not with
PACS.  This result is partly due to the low luminosity of cold 
sources in the PACS wavebands and partly due to the wide range of angular 
resolutions among the five wavebands.

	For our analysis, we consider only those sources which could be 
identified with \gs~in at least two different wavebands with a significance of 
at least 7.  Before fitting a modified black body function to the SED to a source, we additionally stipulated 
that a source must be detected in at least one more waveband with a 
significance of at least 5.  These criteria were satisfied by 224 of the $\sim$800 sources. %Note that the source extractions performed so 
%far, while tested extensively, are preliminary.  

\begin{figure} %figure 5 --old Figure 2 %The number 224 comes from the 13jan21 run.
\centering
%\plotone{ngc7538_11jul22_getsources_3col_v2.eps} 
%\includegraphics[width=5in]{ngc7538_11jul22_getsources_3col_v2.eps} 
%Figure comes from:
%Getsources catalog formatted with ngc7538/getsources/print_maxSED_ellipse.pro
%intermediate greg output files concatenated with cat max70*.txt max160*.txt max250*.txt > dens_overlay_prelim.greg
%Overplotted on column density map with ngc7538/temp_dens/dens_overlay.greg
%\includegraphics[width=5in,angle=-90]{dens_overlay_goodSED.eps}
\includegraphics[width=5in,angle=-90]{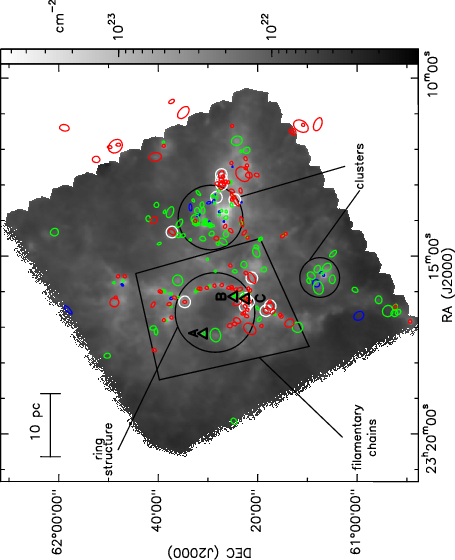}
\caption{ Column density greyscale image overlaid with the 224 compact sources extracted from the 
maps using \gs.  The colored ellipses represent the peak wavelength of each source's SED.  Sources that are brightest at 70 \um~or 160 \um~are blue or green, respectively, and sources that are brighter at 250 \um~or longer are red.  White circles indicate the 13 %42 
candidate sites of high-mass star formation highlighted later in Figure \ref{fig:lvsm} and discussed in Section \ref{sec:props}.  The black triangles labeled A, B, and C are the sources discussed in further detail in Section \ref{sec:props}. The large 
group of blue sources just to the west of center are embedded 
within the large dusty complex containing the optically bright \ion{H}{2} 
region which is the region's most prominent visible feature.  Red sources that are outside the column density map are in the area mapped by SPIRE but not by PACS. \label{fig:threecolsources} }

%\caption{Three-colour image of the compact sources extracted from the 
%maps using \gs, covering the same area as Figure~\ref{fig:threecol}.  
%As in Figure~\ref{fig:threecol}, the three wavelengths represented are 
%SPIRE  250 \um~(\emph{red}), PACS 
%160 \um~(\emph{green}), and PACS 70 \um~(\emph{blue}).  The large 
%group of blue sources just to the west of center are embedded 
%within the large dusty complex containing the optically bright \ion{H}{2} 
%region which is the region's most prominent visible feature.  In producing 
%this image, the relative fluxes of the different bands have been adjusted 
%for visual clarity, but all sources within a band are coloured according 
%to the same scale.  Grey circles indicate the same 42 candidate sites of 
%high-mass star formation highlighted later in 
%Figure \ref{fig:lvsm}.\label{fig:threecolsources}} 
\end{figure}

	Figure~\ref{fig:threecolsources} shows the molecular hydrogen column density image overlaid with the  
sources extracted by \gs.  The image shows that the compact 
sources divide into two groups: tight clusters of point-like sources which 
are bright at 70~\um~(blue and green in Figure \ref{fig:threecolsources}) and long, 
filamentary chains of sources which are typically brighter at the longer 
wavelengths (red in Figure \ref{fig:threecolsources}).  
 Observations by \citet{schn2012} support the idea that the most massive YSOs preferentially form in clusters at the junction of filaments.
 %\citet{schn2012} show that cooler, less-evolved sources lie 
%along filamentary structures while hotter, more evolved sources lie in clusters at 
%the intersections of filaments.  
However, the observed distribution of sources in this region may be due more  %more likely associated with 
to their proximity to high-mass versus low-mass sites of star formation.

%This division suggests that the cooler, less-evolved sources lie 
%along filaments while the hotter, more evolved sources lie in clusters at 
%the intersections of filaments as seen by \citet{schn2012}.  Some of the reddest filaments are arranged 
%in arc-like structures, suggesting that star formation has been triggered along shock 
%fronts or in material swept up by winds or \ion{H}{2} regions.

\begin{figure} %Figure 6 --old Figure 4
\begin{center}
\includegraphics[width=5 in,height=6 in,angle=0]{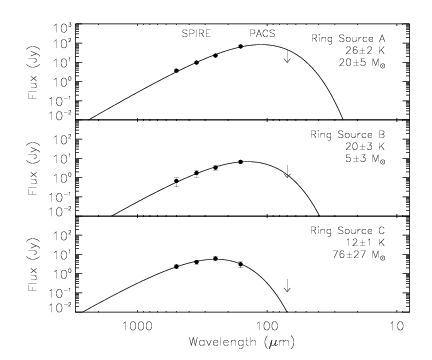} 
\caption{Spectral energy distributions for the three %of the 
sources around the 
edge of the ring in eastern NGC~7538 that are discussed in Section \ref{sec:getsources}.  The points show the integrated fluxes of 
each source as measured by \gs~ after the application of the flux 
scaling discussed in Section \ref{sec:props}.  The lines show the best-fitting 
single-temperature grey body of the form shown in Eqn.~\ref{eq:gb}.  The source 
labels--A, B, and C--correspond to the sources marked in   
Figs.~\ref{fig:ring} and \ref{fig:threecolsources}.  The temperature and mass of each 
source derived from the grey body fit to the four longest wavelengths as discussed in Section \ref{sec:props} are indicated. \label{fig:sedfig}} %As discussed in Sec~\ref{sec:props}, the
% 70~\um~data were not included in the SED fitting, but are included in this figure. \label{fig:sedfig}} 
\end{center}
\end{figure}

	Figure~\ref{fig:sedfig} shows the SEDs of three %the three {\color{red}stage?} sources circled 
%in Figure~\ref{fig:ring}.  {\color{red}Need to show in Fig 1 or 2}
 sources identified in NGC~7538 and labeled with triangles in Figure \ref{fig:threecolsources}.  These sources are a representative sample of the
sources which define the ring structure in eastern NGC~7538 discussed in Section \ref{sec:ring} %(see Figure \ref{fig:ring}; discussed further in Sect.~\ref{sec:ring}) 
 and are well-fit by single-temperature 
grey body SEDs.  Source~A  ($\alpha,\delta$ (J2000) = 23:17:10.3, +61:31:14) is luminous ($L = 600$~\lsun) and extended 
(diameter $\simeq$ 0.6~pc).  We estimate its mass to be 20$\pm$5~\msun, 
which, combined with its high temperature of 26$\pm$2~K, suggests it 
might be forming a high-mass star or a small cluster.  Source~B (at $\alpha,\delta$ (J2000) = 23:16:07.5, +61:25:14) is a more 
typical low-mass core, with a luminosity and diameter of 38~\lsun~and 0.4~pc, respectively.  Having 
a relatively high temperature of 20$\pm$3~K, it is 
likely already undergoing star formation. 
Source~C (at $\alpha,\delta$ (J2000) = 23:16:10.8, +61:22:52) is one of the candidate cold, 
high-mass clumps identified %found
 in this study and is listed in Table \ref{tab:hmdc} as HMDC 8.  With a luminosity and diameter of
18~\lsun~and 0.4~pc, respectively, it has a mass of 76$\pm$27~\msun, and a 
temperature of 12$\pm$1~K. 
%%%%%Source A,B,C from original version:
%Source~A is luminous ($L = 700$~\lsun) and extended 
%(diameter $\simeq$ 0.5~pc).  We estimate its mass to be 13$\pm$1~\msun, 
%which, combined with its high temperature of 30.4$\pm$0.9~K, suggests it 
%might be forming a massive star or a small cluster.  Source~B is a more 
%typical low-mass core, with a luminosity and diameter of 7~\lsun~and 0.5~pc, respectively.  Having 
%a relatively high temperature of 26.1$\pm$0.9~K, it is 
%likely already undergoing star formation. 
%Source~C is one of the candidate cold, 
%high-mass cores identified %found
% in this study.  With a luminosity and diameter of
%21~\lsun~and 0.2~pc, respectively, it has a mass of 40$\pm$20~\msun, and a 
%temperature of 14.2$\pm$0.9~K. 
%
 Thus, it is within the $L/M < 
1$~\lsun/\msun~regime where unevolved high-mass sources are thought to lie (see Section \ref{sec:props} below).  Note 
that because it is cold, Source~C is barely visible in the 70~\um~image (see Figure \ref{fig:ring})
but becomes much brighter at longer wavelengths.  %\emph{Herschel}'s ability to 
%both detect and characterize cold, high-mass sources like this one make it an especially exciting tool for the study of high-mass star 
%formation in nearby regions.

%Figure produced with ~/ngc7538/analysis/seds/sed_fit_getsources_v13jan21test.pro
%Original figure ~/ngc7538/analysis/seds/13jan21_test/13jan21_testsed_fig.eps then eps file copied locally.

\section{Discussion}
\subsection{The Ring}
%see HIA notebook calculation page 74. (using go flux and @masses_calculate)
We determine an order-of-magnitude estimate of 500 M$_{\odot}$ for the mass of the ring by assuming a temperature of 15~K and converting the observed 250~\um~ flux in the ring %and converting to a dust mass of 500
 based on a dust opacity at 250~\um~of 2 as given by \citet{h83}.  
 Using a simple order-of-magnitude energy calculation, we determine that the energy required to move that mass from a centrally concentrated sphere out to a spherical shell of radius 8~pc based on an assumed expansion rate of 1~km~s$^{-1}$  would be on the order of 10$^{45}$~erg.  While the observations suggest that a two dimensional geometry may be more appropriate, we choose spherical symmetry for this simple illustrative first-order approximation.  The assumed expansion rate of 1~km~s$^{-1}$ is typical for a supernova remnant near the end of the radiative expansion %snowplow
 phase when the expansion rate becomes that of the sound speed of the ambient medium.  This energy is several orders of magnitude lower than the $\sim10^{51}$~erg released in a typical supernova collapse of a massive star \citep{wils1985}.  %on the order of 10^51 erg
It may instead be similar to the amount of energy contributed by the stellar wind of a massive star, but we see no indication for the presence of such a source within the ring.  We also look into the possibilities that a runaway O star that may have originated from within the ring, or alternatively, that a nearby windy O-star blew out the cavity within a bubble.  %We do not find any O stars within half a degree of the ring.  
The strongest candidate for either of these scenarios is HIP 115424, an O8 star approximately 50\arcmin~($\sim$40~pc) northeast from the center of the ring.  This star is classified as a runaway star with a peculiar tangential velocity of 30~km~s$^{-1}$ \citep{moff1998}.  Despite its relative proximity, the position and kinematics of this star make it unlikely that it originated within the ring.  At its present location, the winds of an O8 star are likely too weak to have had much influence on the ring. %bubble.  
Investigation into the accuracy of the spectral type classification of this star is necessary to explore this possibility further.%provide illumination for this possibility.}

	Based on its geometry, we suggest that this ring might be an 
example of triggered star formation.  Although there is no \ion{H}{2} region associated with this ring as in the case of the triggered star formation in the HOBYS study of N49 \citep{zava2010}, for example, Figures~\ref{fig:ring} and \ref{fig:threecolsources} show that most of the ring is 
%As can be seen in both 
%Figures~\ref{fig:threecolsources} and \ref{fig:ring}, both sides of the ring are 
delineated by compact sources and cool dust and show that it is a coherent velocity structure.  Our SED fits show 
that very few of these sources have temperatures exceeding 30~K and that 
many are somewhat colder.  Several of our massive cold %Stage E 
 clumps---sites with the potential for intermediate- to high-mass star formation---lie along 
the ring and especially in the cluster of sources to the south.  
\citet{fries07} also located several candidate sites of early high-mass star 
formation in this cluster.  This cluster of sources make up the Infrared Dark Cloud (IRDC) 
G111.80+0.58 \citep{fries08}.  Our data reveal that this IRDC is not merely 
part of a filament, but actually at the intersection of a ring-like %bubble-like 
structure with surrounding filaments.

\subsection{Source Properties and Energetics}
\label{sec:props}

	Given the wide range of angular resolutions, and the likely presence of thermal
substructure within each source, each waveband is sensitive to slightly 
different physical components within each source.  
 For example, the warmer and more compact central regions of protostellar objects would be more prominent at shorter wavelengths within the smallest beams, while longer wavelengths with their larger beams are more sensitive to their cooler and more extended envelopes.
%For example, at shorter wavelengths, 
%we see the more compact, likely warmer central regions of protostellar objects, 
%while at longer wavelengths we see their more extended, cooler envelopes.  
To account for this size-wavelength dependence, we adopt the scaling prescription 
of \citet{hobys}, which was elaborated further by \citet{luong}.   In general, this technique is applied to the regions in the HOBYS program because of the large distances involved, but it is unnecessary for resolved cores in the nearby low-mass star formation clouds. According to this 
prescription, we set the deconvolved source size using a fit to the 160~\um~image and 
then scale the 250~\um, 350~\um, and 500~\um~fluxes down according to

\begin{equation}
S_{{\rm int},\lambda}^{\rm scaled} (< r_{\lambda}) = S_{{\rm 
int},\lambda}^{\rm original} \left( \frac{r_{160~\mu m}}{r_{\lambda}} \right), \label{eq:scale}
\end{equation}

\noindent where $r_{\lambda}$ is the deconvolved radius of the source %mean of the semi-major and semi-minor 
%axes of the ellipse 
over which the flux, $S_{{\rm 
int},\lambda}$, is integrated.  This flux scaling method assumes that the source size obtained in the 160 micron image is accurate, that the emission at the longer wavelengths (250~\um, 350~\um, and 500~\um) is optically thin, and that the flux varies linearly with angular radius.  For %high-mass 
dense cores and clumps, it is reasonable to assume that they are optically thin at these wavelengths.  Likewise, the linear variation of the flux is a good approximation, %probably reasonable, 
at least until the point that an H{\sc ii} region forms.  Although testing non-linear models is beyond the scope of this paper, if the flux variation does not vary linearly with angular radius, it is likely that the temperature would be overestimated and consequently the mass would be underestimated. %See Quang's paper for elucidation.

	After scaling the source fluxes in this way, we find that 224 %The number of sources that have Fit quality = good in sed_fit_getsources_no70
sources out of the nearly 800 preliminary sources %817
 are well fit %(at the $p = 0.05$  significance level) 
 by a single-temperature grey body SED.  Under the assumption that the data are accurately represented by the given SED function, we define a good fit such that there is a 95\% chance that the calculated $\chi^{2}$  will be less than the $\chi^{2}$ value expected from random variations in the data (i.e. at the p-value $p = 0.05$ statistical significance level.)
 We fit a SED of the form

\begin{equation}  %switched to a reference wavelength of kappa=.1 at 300 micron
S(\nu) = 
\frac{M_{\rm gas+dust}\kappa_{300}}{d^2}\left(\frac{\nu}{\nu_{300}}\right)^{\beta} 
B(\nu,T_{\rm dust})~~,
\label{eq:gb}
\end{equation}

\noindent where $M_{\rm gas+dust}$ is the total mass of the source, $d = 
2.7$~kpc is the distance to the cloud, $\kappa_{300}$ is the the dust 
opacity per unit (gas + dust) mass at 300~\um, $\nu_{300}$ is the frequency 
%of light at
 corresponding to a wavelength of 300~\um, $\beta$ is the dust emissivity index, and 
$B(\nu,T_{dust})$ is the Planck function for dust temperature $T_{\rm 
dust}$.  We adopt a dust opacity of $\kappa_{300}$ = 0.1 cm$^{2}$ g$^{-1}$ 
and a fixed dust emissivity index of $\beta = 2.0$.  These values are 
derived from \citet{h83} and are consistent with those used in other HOBYS studies (e.g., \citet{hobys}) as well as other SPIRE 
Galactic key programmes. %, including other HOBYS studies (e.g., \citet{hobys}).  
Note, however, that the 
dust opacity and emissivity index, $\beta$, are likely to vary with environment \citep{osse1994} % is uncertain, 
which may imply non-systematic uncertainties in our 
measured dust masses %of a factor of two or more 
(e.g., see Sadavoy et al. 2013, submitted). %The median 
%(non-systematic) uncertainty in the fitted temperatures is 16\%.
We assume a gas-to-dust ratio of 100.  The total amount of mass associated with these 224 sources is on the order of 1$\times$10$^{4}$~\msun.  The overwhelming majority of this mass is contained within sources having temperatures less than 20~K. 
%determined from ~/ngc7538/temp_dens/tempdist.pro
% temperature cutoff (K) =        20.000000
%Total mass (Msun) =        13009.102
%Total mass above cutoff Temp; fraction of total =        1704.7063      0.13103951
%Total mass below cutoff Temp; fraction of total =        11304.396      0.86896049

 We do not include the 70~\um~data in the SED fitting (see \citet{hill2012} for further discussion).  At wavelengths shorter than 100~\um, the emission may require an additional temperature component to properly include the contribution from the warmer material associated with the protostar.  Also, the 70~\um~emission may arise from very small dust grains (VSG) thereby changing the dust opacity law in this regime.  Thus, in cases where the inner warm envelope material or VSG emission 
dominate the 70~\um~flux associated with the bulk envelope (the material 
primarily traced by the $\lambda \ge$ 160~\um~bands), the observed 70~\um~emission is typically higher than the SED fits at that wavelength. %is an upper limit. 

	With flux scaling applied to remove some of the influence of 
each source's cold outer envelope, the SED %fits improve and, as expected, 
%the source temperatures rise by several degrees.  These 
fits are more 
representative of the interior parts of the sources where star formation, 
if present, would occur.  For the purposes of assessing the 
evolutionary states of these sources, the conditions in their interiors are 
more important than those in their envelopes.  For example, a cool interior is a more indicative %truer 
sign of youth than a cool envelope.

	In the future, we could leverage the information about each source's 
emission on multiple spatial scales to construct more sophisticated, 
multi-temperature models, but for the present purpose of developing an 
overall picture of the star-formation activity in NGC~7538, we maintain 
this simple approach.  Note that this procedure underestimates the 
source masses because it excludes flux from the outer envelope of each source as well as flux from wavelengths longer and shorter than those observed with \emph{Herschel}.  %This is an acceptable risk when 
We deem this approach %risk 
to be acceptable for our purposes since we are trying to identify conservatively sites %make conservative identifications of sites 
of high-mass star formation.  

	Having derived each identified source's temperature and %dust 
mass  %(assuming a gas-to-dust ratio of 100),
from Equation~\ref{eq:gb}, we compute their \emph{Herschel} grey body luminosities as:

\begin{equation}
L = 4 \pi d^2 \int_{0}^{\infty} S(\nu) d\nu.
\end{equation}

\begin{figure} %Figure 7 --old Figure 3
\begin{center}
\includegraphics[width=6in]{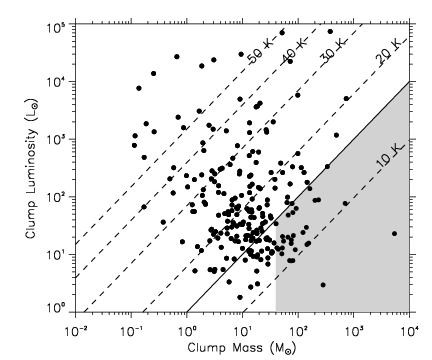}
\caption{ Herschel grey body luminosity vs.~total mass (dust + gas) for compact sources in 
NGC~7538.  
Only the 224 sources whose SEDs were well-fit by a grey body function with 
$\beta = 
2$ are included in the plot.  The diagonal dashed lines represent loci of 
constant temperature, with the respective temperatures indicated on each 
line.  The 
thick solid line represents $L/M$ = 1 \lsun/\msun, equivalent to a 
temperature of 
about 
15.7~K.  The shaded region (M $\geq$ %20
40~\msun~and 
L/M $\leq$ 1 \lsun/\msun) highlights massive cold clumps which we identify 
as 
potential 
precursors of high-mass stars. 
\label{fig:lvsm}}
\end{center}
\end{figure}

\noindent In 
Figure~\ref{fig:lvsm}, we plot the luminosity of each source versus its 
total mass.  The luminosity plotted here is that %the dust luminosity 
derived 
from the SED which, for cool, starless sources, should be similar to its 
bolometric luminosity.  By taking the ratio of each source's luminosity to its mass, we 
can assess the degree to which it is affected by internal heating, and 
therefore the likelihood that it is already forming stars \citep[e.g.][]{andr2008, hobys}.  

	The evolutionary state of a low-mass protostellar core is typically defined 
according to a scheme of classes, between Class~0 and Class~III, using a 
combination of infrared and submillimeter observations \citep[e.g.][]{lada1987,andre}.  This 
technique is generally less applicable to high-mass star formation for several reasons.  
First, the technique relies on positional coincidence to establish whether or not an
 infrared source is embedded within a given molecular cloud core.  
True physical association of infrared sources with dusty clumps is 
difficult to establish in high-mass star-forming regions, however, because these 
regions are typically distant and in clustered environments.  
 Second, if a low-mass core contains an infrared point source, it is reasonable to assume that star formation is already underway and the star being formed is likely going to be the largest in the system.  In a high-mass core, however, the high-mass star may not be the first star in the cluster to form.  The presence of an infrared point source in a high-mass core may result from one or several low-mass protostars that formed before the most massive star in the cluster, and does not necessarily correspond to what will become the most massive star in the cluster.

%The presence of an infrared point source within a low-mass core 
%would automatically advance the classification of the core further along the evolutionary sequence.  
%The presence of an infrared point source in a high-mass core does 
%not, however, preclude the possibility that it may still form a high-mass star because 
%such stars may form among clusters of low-mass stars.   In fact, some 
%models of massive star formation might predict that the presence of an 
%embedded infrared source in a massive clump would \emph{increase} the odds 
%of that clump forming a massive star, simply by proving that parts of the 
%clump are gravitationally bound.

	For the above reasons, we employ the mass-luminosity diagram as a tool 
for characterizing the evolutionary states of the compact sources in 
NGC~7538.  The mass-luminosity diagram helps separate sources according to 
their energetics and can be useful for separating Class I YSOs from Class 0 protostars and prestellar cores in the low mass regime and infrared-bright massive YSOs from infrared-quiet protostellar objects and starless dense cores in the high mass regime % identifying pre-stellar source evolutionary sequences
 \citep{bon96,mol08,hennemann,roy11}.  Due to their 
high opacities, unevolved dusty clumps with no internal heat source ought to 
be a few degrees colder than the ambient temperature of their parent molecular clouds, 
having temperatures of about 15~K and luminosity-to-mass ratios $L/M 
\lesssim$ 1~\lsun/\msun~ \citep{roy11}.  Indeed, Roy et al.~divided high-mass sources in Cygnus~X 
into two evolutionary categories: ``Stage E,'' denoting 
externally-heated sources, and ``Stage A,'' denoting sources at or above 
the ambient temperature due to heating by accretion.  Stage E overlaps with 
the prestellar and Class 0 stages in the low-mass paradigm.  In the 
mass-luminosity diagram, Stage E sources lie below $L/M \sim 
1$~\lsun/\msun, while Stage A sources lie strictly above $L/M = 
1$~\lsun/\msun.

	To the Stage E classification, we add the additional requirement 
that a source must have a mass of at least %20
40~\msun. %to be considered capable of forming a high-mass star.  
%This assumption is reasonable, given 
%a star formation efficiency of 30\% or greater \citep{lada}, and is in 
%keeping with the criterion used by \citet{m07,hobys} in their studies of 
%Cygnus~X and several other HOBYS fields (Rosette and RCW~120).  
 Given a star formation efficiency of 30\% or greater \citep{lada}, a core would require on the order of 20~\msun~to form a high-mass star.  In their study of Cygnus~X, \citet{m07} use a criterion of 40~\msun~as the mass required for a core to form a high-mass star.  In the study of several other HOBYS fields (Rosette, W48 and RCW 120), \citet{hobys} and \citet{luong} use a lower criterion of 20~\msun~to enlarge the census to cores able to form an intermediate-mass star.
% In their studies of Cygnus~X and several other HOBYS fields (Rosette and RCW~120) \citet{m07,hobys} use a criterion of 20~\msun~as the mass required for a core to form a high-mass star.  
Most of those objects are more compact (on the order of 0.1~pc compared to $\sim$0.7~pc for the objects in this study), however, so we choose a more conservative lower  limit of 40~\msun~in order to isolate the highest mass objects in the 
NGC~7538 region.

	We find 27 %42 
compact objects in NGC~7538 that satisfy the 
criteria $M_{\rm gas+dust} \geq %20
40$~\msun~and $L/M~\leq~1$~\lsun/\msun.  
These high-mass precursor candidates  %objects 
occupy the shaded region in Figure~\ref{fig:lvsm}.  Applying the further criteria that a source must be located within the region mapped by both PACS and SPIRE, that its temperature must be greater than 10~K, and that it be undetected at 24~\um, 13 sources remain. A list of properties for these 13 sources
is given in Table \ref{tab:hmdc}.  
%This list is preliminary as the sources given here were only detected with one source extraction algorithm.  
 As source extraction techniques evolve, this initial list of 
sources will need to be confirmed with a second detection method. 
These 13 %42  
sources range in 
%%%%%%%%%%%%%%%%%%%%%%%%%%%%%%%%%%%%%%%%%%%%%%%%%%%%
%The following values come from running laptop:~/ngc7538/analysis/seds/13febxx_scaled_no70/hmpo_list_tex.pro
%%%%%%%%%%%%%%%%%%%%%%%%%%%%%%%%%%%%%%%%%%%%%%%%%%%%
diameter from {\textbf 0.4~pc to 1.1~pc} %(before deconvolution) 0.4~pc to 2.5~pc %2.6 pc, %0.18~pc to 2.4~pc, %0.41 to two sigdigs
in mass from {\textbf 40~\msun~to a few hundred solar masses, } %(before deconvolution) 44~\msun~to %23~\msun~to 
%$\sim$1000~\msun, 
and in density from $4\times10^{3}$~cm$^{-3}$ to $4\times10^{4}$~cm$^{-3}$.  %$1\times10^{3}$~cm$^{-3}$ to $1\times10^{6}$~cm$^{-3}$. 
%$5\times10^{2}$~cm$^{-3}$ to $1\times10^{6}$~cm$^{-3}$.  
Their median diameters, masses, and densities 
are {\textbf 0.7~pc, 80~\msun, and $9\times10^{3}$~cm$^{-3}$} %(for un-deconvolved sizes) 0.7~pc, 70~\msun, and $2\times10^{4}$~cm$^{-3}$ %50~\msun, and $1\times10^{4}$~cm$^{-3}$ %0.6~pc, 300~\msun, and $1\times10^{5}$~cm$^{-3}$ %61to two sigdigs
respectively.  We 
define the deconvolved diameter of a source using the major and minor axes measured at the %the length of its major axis at the 
shortest wavelength at which the source was resolved (which is 160~\um~in all 13 cases.  Its volume is defined as 
that of an ellipsoid having the same major and minor axes as the source, plus 
a third axis whose length is the arithmetic mean of the lengths of the other two.

Higher resolution data may reveal fragmentation within some of these sources, but we still present them here as a first cut of the potential intermediate and high mass star forming sites in NGC~7538.  While many of these sites don't fulfill the criteria set by \citet{kauf2010}, they still appear to be cold clumps of significant mass and are worthy of further investigation.

\begin{table}[h]
\begin{center}
\caption{Properties of the 13  high-mass dense clump (HMDC) candidate objects.  \label{tab:hmdc}}
%\centering
\begin{tabular}{|l|l|l|r|c|r|r|r|}
\hline\hline
HMDC   & RA       & Decl.    & Mass             & L/M\tablenotemark{a}  & Size\tablenotemark{a}  & Temp\tablenotemark{a} & Density\tablenotemark{a} \\ 
               & h:m:s    & d:m:s    & \msun           & \lsun/\msun  & pc      & K         & cm$^{-3}$ \\ 
\hline
%ordered by RA:
%with deconvolved sizes:
1 & 23:12:42.5 & 61:27:47.9 & 200 & 0.43 & 0.9 & 13 & 8 $\times$10$^{3}$ \\ %425
2 & 23:12:51 & 61:27:46.7 & 80 & 0.66 & 0.7 & 14 & 8 $\times$10$^{3}$\\ %393
3 & 23:12:58.9 & 61:27:44.1 & 150 & 1.2 & 0.8 & 15 & 9 $\times$10$^{3}$\\ %287
%1 & 23:13:20.4 & 61:25:25.8 & 63 & 1.25 & 0 & 15.4\\ %88 %added after deconvolution prob'ly has 8, 24 mu emission
4 & 23:13:20.7 & 61:28:47.7 & 40 & 0.85 & 0.4 & 14 & 2 $\times$10$^{4}$\\ %408 %added after deconvolution
5 & 23:13:25 & 61:25:6.6 & 340 & 0.99 & 1.0 & 15 & 7 $\times$10$^{3}$\\ %325
6 & 23:14:19.5 & 61:37:45.4 & 82 & 0.26 & 0.8 & 12 & 5 $\times$10$^{3}$\\ %450
7 & 23:15:36.4 & 61:21:38.9 & 140 & 0.21 & 1.1 & 11 & 4 $\times$10$^{3}$\\ %483
%3 & 23:15:53.7 & 61:30:41.9 & 44 & 0.28 & 0 & 12\\ %200 %added after deconvolution; has 24 mu emission, 
%2 & 23:15:57.6 & 61:21:58.4 & 52 & 0.25 & 0 & 11.7\\ %152 %added after deconvolution; has 8, 24 mu emission,
8\tablenotemark{b}& 23:16:10.8 & 61:22:52.4 & 76 & 0.24 & 0.4 & 12 & 4 $\times$10$^{4}$\\ %232
9 & 23:16:16.8 & 61:22:16.9 & 160 & 0.10 & 0.5 & 10 & 4 $\times$10$^{4}$\\ %346
10 & 23:16:17.6 & 61:35:7.1 & 70 & 0.11 & 0.5 & 10 & 2 $\times$10$^{4}$\\ %329
11 & 23:16:23 & 61:17:55.1 & 75 & 0.21 & 0.6 & 11 & 1 $\times$10$^{4}$\\ %357
12 & 23:16:24.3 & 61:22:56.2 & 42 & 0.31 & 0.5 & 12 & 1 $\times$10$^{4}$\\ %399
13 & 23:16:31.3 & 61:18:46.9 & 81 & 0.10 & 1.1 & 10 & 6 $\times$10$^{3}$\\ %465

%without deconvolved sizes:
%%1 & 23:12:27.4 & 61:22:38 & 41 & 0.39 & 0.5 & 13 & 2 $\times$10$^{4}$\\ %188 has 24 mu emis. nearby, no 8 mu obs.
%1 & 23:12:42.5 & 61:27:48 & 100 & 0.76 & 1.7 & 14 & 4 $\times$10$^{3}$\\ %425
%2 & 23:12:51.0 & 61:27:47 & 66 & 0.80 & 0.8 & 14 & 7 $\times$10$^{3}$\\ %393
%%4 & 23:12:54.7 & 61:26:56 & 74 & 0.83 & 0.5 & 14 & 3 $\times$10$^{4}$\\ %79 has 24 micron emission, 8 mu ok
%3 & 23:12:58.9 & 61:27:44 & 180 & 1.0 & 1.2 & 15 & 1 $\times$10$^{4}$\\ %287
%4 & 23:13:25.0 & 61:25:07 & 970 & 0.36 & 2.5 & 13 & 2 $\times$10$^{4}$\\ %325
%5 & 23:14:19.5 & 61:37:45 & 59 & 0.34 & 0.9 & 12 & 4 $\times$10$^{3}$\\ %450
%6 & 23:15:36.4 & 61:21:39 & 96 & 0.28 & 1.3 & 12 & 3 $\times$10$^{3}$\\ %483
%7\tablenotemark{a} & 23:16:10.8 & 61:22:52 & 52 & 0.30 & 0.4 & 12 & 3 $\times$10$^{4}$\\ %232
%%10 & 23:16:12.9 & 61:22:32 & 48 & 0.96 & 0.5 & 15 & 2 $\times$10$^{4}$\\ %44  has 24 mu emission, 8 mu ok
%8 & 23:16:16.8 & 61:22:17 & 160 & 0.10 & 0.7 & 10 & 4 $\times$10$^{4}$\\ %346
%9 & 23:16:17.6 & 61:35:07 & 63 & 0.12 & 0.6 & 10 & 2 $\times$10$^{4}$\\ %329
%10 & 23:16:23.0 & 61:17:55 & 75 & 0.21 & 0.7 & 11 & 1 $\times$10$^{4}$\\ %357
%11 & 23:16:24.3 & 61:22:56 & 44 & 0.30 & 0.7 & 12 & 1 $\times$10$^{4}$\\ %399
%%15 & 23:16:29.7 & 61:22:31 & 53 & 0.36 & 0.5 & 12 & 2 $\times$10$^{4}$\\ %171 has 8, 24 micron emission
%12 & 23:16:31.3 & 61:18:47 & 70 & 0.11 & 0.8& 10 & 5 $\times$10$^{3}$\\ %465
%%17 & 23:16:44.3 & 61:37:45 & 47 & 0.16 & 0.4 & 11 & 3 $\times$10$^{4}$\\ %102 has 24 mu emission, no 8 mu obs.
%%%%%%%%%%%%%%-----------------begin deluxetable
\hline
\end{tabular}
\tablenotetext{a}{~luminosity/mass; clump diameter; temperature; volume density}
\tablenotetext{b}{~This is source `C' indicated in Figs.~\ref{fig:ring} and \ref{fig:threecolsources} and discussed in Sec.~\ref{sec:getsources}.}
\end{center}
\end{table}

	These 13 %42 
high-mass Stage E sources are distributed spatially throughout NGC~7538.  
They are generally visible as the reddest objects in 
Figure~\ref{fig:threecolsources} and are highlighted with white circles in that same figure.  A more detailed analysis of the cloud 
structure and source distribution in NGC~7538 is beyond the scope of this first-look paper, but would be interesting to look into in a followup study. %will be presented in a forthcoming paper.

\section{Conclusions}

	We have reported the first results of the \emph{Herschel} HOBYS 
observations of the nearby high-mass star-forming region, NGC~7538.  The 
thermal dust emission shows many compact sources distributed along 
filaments.  We have detected nearly 800 compact sources and characterized the SEDs 
of 224 of them.  Of these latter sources, we identify 13 as high-mass dense clump candidates, %42 
 potential sites of 
future intermediate- to high-mass star formation.  We present the characteristics of these select high-mass dense clump objects which require further follow-up observations to confirm that star formation is underway and determine the source kinematics. 

We also report the discovery of a ring of cool thermal dust emission of as-yet unknown origin.  With additional data from the JCMT, we further characterize the ring and determine properties of the ring such as its extent and energetics.  We look into several possible origin scenarios for the ring, none of which provide a satisfactory explanation.  %that rule out its origin from a supernova explosion.  
We detect a large number of cold sources along the ring's filamentary edge.  
%This ring may represent an example of ongoing triggered star formation. % in progress.

\acknowledgements
We acknowledge the support of the Canadian Space Agency (CSA) via a Space Science Enhancement Program grant, the National Science and Engineering Rsearch Council (NSERC) via a Discovery grant, and the National Research Council of Canada (NRC).

SPIRE has been developed by a consortium of institutes led by Cardiff 
University (UK) and including: Univ. Lethbridge (Canada); NAOC (China); 
CEA, LAM (France); IFSI, Univ. Padua (Italy); IAC (Spain); Stockholm 
Observatory (Sweden); Imperial College London, RAL, UCL-MSSL, UKATC, Univ. 
Sussex (UK); and Caltech, JPL, NHSC, Univ. Colorado (USA). This 
development has been supported by national funding agencies: CSA (Canada); 
NAOC (China); CEA, CNES, CNRS (France); ASI (Italy); MCINN (Spain); SNSB 
(Sweden); STFC, UKSA (UK); and NASA (USA).

PACS has been developed by a consortium of institutes led by MPE (Germany) 
and including UVIE (Austria); KU Leuven, CSL, IMEC (Belgium); CEA, LAM 
(France); MPIA (Germany); INAF-IFSI/OAA/OAP/OAT, LENS, SISSA (Italy); IAC 
(Spain). This development has been supported by the funding agencies BMVIT 
(Austria), ESA-PRODEX (Belgium), CEA/CNES (France), DLR (Germany), 
ASI/INAF (Italy), and CICYT/MCYT (Spain).

The James Clerk Maxwell Telescope is operated by the Joint Astronomy Centre on behalf of the Science and Technology Facilities Council of the United Kingdom, the National Research Council of Canada, and (until 31 March 2013) the Netherlands Organisation for Scientific Research.

This research has made use of the SIMBAD database, operated at CDS, Strasbourg, France

KLJR acknowledges support by the Agencia Spaziale Italiana (ASI)
under contract number I/005/11/0.

We would also like to thank the anonymous referee(s) for his/her/their comments which have significantly improved this manuscript.


%% in parentheses, after the keyword, but they will not be verified.

{\it Facilities:} \facility{Herschel, JCMT}


%\appendix
%\section{Column Density and Temperature Maps}

%% The following command ends your manuscript. LaTeX will ignore any text


\begin{thebibliography}{}
\bibitem[Anderson et al.(2012)]{ande2012}Anderson, L.~D., et al.~2012, \aap, 542, 10
\bibitem[Andr\'e, Ward-Thompson, \& Barsony(1994)]{andre}Andr\'e, P., Ward-Thompson, D., \& Barsony, M.~1993, \apj, 406, 122
\bibitem[Andr\'e et al.(2008)]{andr2008}Andr\'e, Ph., et al.~2008, \aap, 490, L27
\bibitem[Beaumont \& Williams(2010)]{beau2010}Beaumont, C.~N., \& Williams, J.~P.~2010, \apj, 709, 791
\bibitem[Bernard et al.(2010)]{bern2010}Bernard, J.-P., et al.~2010, \aap, 518, L88
%\bibitem[Beuther et al.(2007)]{beut2007}Beuther, H., Churchwell, E.~B., McKee, C.~F., \& Tan, J.~C.~2007, Protostars and Planets V, 165
\bibitem[Bontemps et al.(1996)]{bon96}Bontemps, S., Andr\'e, P., Terebey, S., \& Cabrit, S.~1996, \aap, 311, 858
\bibitem[Churchwell et al.(2006)]{chur2006}Churchwell, E., et al.~2006, \apj, 649, 759
\bibitem[Crampton, Georgelin, \& Georgelin(1978)]{cgg78}Crampton, D., Georgelin, Y.~M., \& Georgelin, Y.~P.~1978, \aap, 66, 1
\bibitem[Davis et al.(1998)]{davis1998}Davis, C.~J., Moriarty-Schieven, G., Eisl{\"o}ffel, J., Hoare, M.~G., \& Ray, T.~P.~1998, \aj, 115, 1118
\bibitem[Frieswijk et al.(2007)]{fries07}Frieswijk, W.~W.~F., Spaans, M., Shipman, R.~F., Teyssier, D., \& Hily-Blant, P.~2007, \aap, 475, 263
\bibitem[Frieswijk et al.(2008)]{fries08}Frieswijk, W.~F., et al.~2008, \apj, 685, L51
\bibitem[Griffin et al.(2010)]{spire}Griffin, M., et al.~2010, \aap, 518, L3
\bibitem[Hennemann et al.(2010)]{hennemann}Hennemann, M., et al.~2010, \aap, 518, L84
\bibitem[Hennemann et al.(2012)]{henn2012}Hennemann, M., et al.~2012, \aap, 543, L3
\bibitem[Hildebrand(1983)]{h83}Hildebrand, R.~H.~1983, \qjras, 24, 267
\bibitem[Hill et al.(2011)]{hill2011}Hill, T., et al.~2011, \aap, 533, 94
\bibitem[Hill et al.(2012)]{hill2012}Hill, T., et al.~2012, \aap, 542, 114
\bibitem[Kauffmann \& Pillai (2010)]{kauf2010}Kauffmann,~J., \& Pillai,~T.~2010, \apj, 723, L7
\bibitem[Lada (1987)]{lada1987}Lada, C.~J.~1987, IAUS, 115, 1
\bibitem[Lada \& Lada(2003)]{lada}Lada, C.~J., \& Lada, E.~A.~2003, \araa, 41, 57
\bibitem[Men'shchikov et al.(2010)]{gs1}Men'shchikov, A., et al.~2010, \aap, 518, L103
\bibitem[Men'shchikov et al.(2012)]{gs2}Men'shchikov, A., et al.~2012, \aap, 542, 81
\bibitem[Minier et al.(2013)]{mini2013}Minier, V., et al.~2013,  \aap, 550, 50
\bibitem[Moffat et al.(1998)]{moff1998}Moffat, A.~F.~J., et al.~1998, \aap, 331, 949
\bibitem[Molinari et al.(2008)]{mol08}Molinari, S., et al.~2008, \aap, 481, 345
\bibitem[Moscadelli et al.(2009)]{mos09}Moscadelli L., et al.~2009, \apj, 693, 406
\bibitem[Motte, Andr\'e, \& Neri(1998)]{man98}Motte, F., Andr\'e, P., \& Neri, R.~1998, \aap, 336, 150
\bibitem[Motte et al.(2001)]{m01}Motte, F., et al.~2001, \aap, 372, L41
\bibitem[Motte et al.(2007)]{m07}Motte, F., Bontemps, S., Schilke, P., Schneider, N., Menten, K.~M., \& Brogui\`ere, D.~2007, \aap, 476, 1243
\bibitem[Motte et al.(2010)]{hobys}Motte, F., et al.~2010, \aap, 518, L77
\bibitem[Nguyen Luong et al.(2011)]{luong}Nguyen Luong, Q.,  et al.~2011, \aap, 535, 76
\bibitem[Ossenkopf  \& Henning (1994)]{osse1994}Ossenkopf, V., \& Henning, T.~1994, \aap, 291, 943
\bibitem[Pickett et al.(1998)]{pick98}Pickett, H.~M., et al.~1998, JQSRT, 60, 883
\bibitem[Pilbratt et al.(2010)]{pilb}Pilbratt, G.~L., et al.~2010, \aap, 518, L1
\bibitem[Poglitsch et al.(2010)]{pacs}Poglitsch, A., et al.~2010, \aap, 518, L2
\bibitem[Reed(2005)]{reed}Reed, B.~C.~2005, \aj, 130, 165
\bibitem[Reid \& Wilson(2005)]{rw05}Reid, M.~A.~\& Wilson, C.~D.~2005, \apj, 625, 891
\bibitem[Rivera-Ingraham et al.(2013)]{rive2013} Rivera-Ingraham, A., et al.~2013, \apj, 766, 85
\bibitem[Roussel (2012)]{rous2012}Roussel, H. 2012, arXiv:1205.2576v1
\bibitem[Roy et al.(2011)]{roy11}Roy, A., et al.~2011, \apj, 727, 114
\bibitem[Schneider et al.(2010)]{schn2010} Schneider, N., et al.~2010, \aap, 518, 83
\bibitem[Schneider et al.(2012)]{schn2012} Schneider, N., et al.~2012, \aap, 540, L11
\bibitem[Ungerechts, Umbanhowar, \& Thaddeus(2000)] {unge2000}Ungerechts, H., Umbanhowar, P., \& Thaddeus, P.~2000, \apj, 537, 221
\bibitem[Wilson(1985)]{wils1985} Wilson, J.~R.~1985, in Numerical Astrophysics, ed. J.~M.~Centrella, J.~M.~Leblanc, \& R.~L.~Bowers (Boston: Jones \& Bartlett), 422
\bibitem[Wynn-Williams, Becklin, \& Neugebauer(1974)]{ww74}Wynn-Williams, C.~G., Becklin, E.~E., \& Neugebauer, G.~1974, \apj, 187, 473
\bibitem[Zavagno et al.(2010)]{zava2010} Zavagno, A. et al., 2010, \aap, 518, L81
\end{thebibliography}
\end{document}